\documentclass[useAMS,usenatbib, usegraphicx]{mn2e}
\usepackage{epsfig}
\usepackage{amsmath}
\usepackage{amssymb}
\usepackage{natbib}
\newcommand{\kms}{\ensuremath{{\rm km\,s}^{-1}}}
\newcommand{\msun}{\ensuremath{{\rm M}_{\odot}}}

\newcommand{\yr}{\ensuremath{\rm yr}}
\newcommand{\Myr}{\ensuremath{\rm Myr}}
\newcommand{\Gyr}{\ensuremath{\rm Gyr}}
\newcommand{\pc}{\ensuremath{\rm pc}}
\newcommand{\kpc}{\ensuremath{\rm kpc}}

\newcommand{\new}{}
\newcommand{\infinity}{{\infty}}

\newcommand{\rmd}{{\rm d}}

\newcommand{\msmbh}{\ensuremath{M_{\bullet}}}
\title[Recoiled BHs in the Milky Way]{Star clusters around recoiled
black holes in the Milky Way halo}

\author[O'Leary \& Loeb]{Ryan M.\
  O'Leary\thanks{E-mail:roleary@cfa.harvard.edu} and Abraham
  Loeb\thanks{E-mail:aloeb@cfa.harvard.edu}\\Harvard University,
  Department of Astronomy, 60 Garden St., Cambridge, MA 02138, USA\\}

\begin{document}
\maketitle

\begin{abstract}

  Gravitational wave emission by coalescing black holes (BHs) kicks
  the remnant BH with a typical velocity of hundreds of $\kms$. This
  velocity is sufficiently large to remove the remnant BH from a
  low-mass galaxy but is below the escape velocity from the Milky Way
  (MW) galaxy. If central BHs were common in the galactic building
  blocks that merged to make the MW, then numerous BHs that were
  kicked out of low-mass galaxies should be freely floating in the MW
  halo today.  We use a large statistical sample of possible merger
  tree histories for the MW to estimate the expected number of
  recoiled BH remnants present in the MW halo today. We find that
  hundreds of BHs should remain bound to the MW halo after leaving
  their parent low-mass galaxies. Each BH carries a compact cluster of
  old stars that populated the core of its original host galaxy.
  Using the time-dependent Fokker-Planck equation, we find that
  present-day clusters are $\lesssim 1\,$pc in size, and their central
  bright regions should be unresolved in most existing sky
  surveys. These compact systems are distinguishable from globular
  clusters by their internal (Keplerian) velocity dispersion greater
  than one hundred km~s$^{-1}$ and their high mass-to-light ratio
  owing to the central BH.  An observational discovery of this relic
  population of star clusters in the MW halo, would constrain the
  formation history of the MW and the dynamics of BH mergers in the
  early Universe.  A similar population should exist around other
  galaxies and may potentially be detectable in M31 and M33.

\end{abstract}

\begin{keywords}
  galaxies:kinematics and dynamics--galaxies:nuclei--black hole
  physics--gravitational waves--star clusters
\end{keywords}

\section{Introduction}

During the final coalescence of two black holes (BHs), gravitational
waves (GWs) are emitted unisotropically and carry away linear
momentum, thus kicking the merger BH remnant in the opposite direction
\citep{1962PhRv..128.2471P,1973ApJ...183..657B,1983MNRAS.203.1049F}.
The resulting kick velocity of typically hundreds of $\kms$ depends on
the mass ratio of the BHs as well as the spin and orientation of the
binary before coalescence
\citep{2006ApJ...653L..93B,2007PhRvL..98w1102C,2007ApJ...659L...5C,2007PhRvD..76f1502T}. Such
kicks can alter both the the population of nuclear BHs in galactic
bulges \citep{2004ApJ...606L..17M,2006MNRAS.368.1381L,2007ApJ...663L...5V,2007ApJ...667L.133S,2008arXiv0805.1420B} as well
as the core of stars in the bulge itself \citep{2008ApJ...678..780G}.
The discovery of a remnant population of recoiled BHs in the
present-day Universe
can provide a new window into the merger statistics and spin
distribution of the BHs \citep{2006MNRAS.368.1381L,2007ApJ...663L...5V}, as well as test
general relativity in the strong regime.

Recoiling BHs may be detected through their unique flaring
\citep{2008ApJ...676L...5L,2008ApJ...682..758S,2008arXiv0802.3556S}
or observed as spatially and kinematically offset quasars
\citep{2004ApJ...606L..17M}.  If the BH binary was surrounded by
an accretion disk, the ejected remnant BH would carry the disk with it
and shine as a quasar \citep{Loeb}.  A search for spectral shifts in the broad
lines of quasars relative to the narrow emission lines of their parent
galaxies resulted mainly in upper limits \citep{2007ApJ...666L..13B}
and possibly one suggested candidate \citep{2008ApJ...678L..81K} for
an ejected quasar. Unfortunately, these quasars lose their source of
gas for accretion, and have short lifetimes
\citep{Loeb,2008arXiv0805.1420B,2008ApJ...687L..57V}.

Past studies of the observational consequences of GW recoil have
generally focused on massive BHs with a mass $\msmbh \gtrsim
10^7\,\msun$.  Such BHs reside in the most massive galaxies, and
require the greatest and rarest kick velocities in order to escape
their parent galaxies \citep{2007ApJ...662L..63S}.  But since the kick
velocity from BH mergers is independent of the total mass of the
system, the dynamical consequences are most prominent for less massive
BHs which tend to reside in low-mass galaxies with shallower potential
wells and lower escape velocities
\citep{2004ApJ...606L..17M,2005MNRAS.358..913V,2006ApJ...653L..93B,2006MNRAS.368.1381L,2007ApJ...663L...5V,2008arXiv0807.4702T}. In fact, most BHs with
$\msmbh \lesssim 10^5\,\msun$ will likely be kicked out of their
parent galaxies as a result of major mergers
\citep{2006ApJ...653L..93B}. During the hierarchical build-up of the
Milky-Way (MW) galaxy, mergers of low-mass galaxies were common and
should have resulted in a population of freely floating BHs \citep{2005MNRAS.358..913V}.
Since the gravitational potential of an overdense region does not
evolve dramatically during the growth of cosmological structure
\citep{2006astro.ph..3360L} the region that eventually collapsed to
make the MW was able to trap those BHs with the most common kick
velocities ($\lesssim 500~{\rm km~s^{-1}}$) even before the MW had
formed.  Each of the ejected BHs carries with it a star cluster that
used to populate the core of its parent galaxy
\citep{2008ApJ...678..780G, 2008ApJ...678L..81K}.  In this paper we
examine the observational signatures of the ejected star clusters
which are expected to be floating in the MW halo.  The discovery of a
relic population of the star clusters attached to recoiled BHs would
provide a new window for cosmology, allowing one to constrain the
merger history of the MW as well as the formation history of the first
population of massive BHs.

{\new Previous studies have looked at populating the MW halo with
  recoiling BHs as analysed again here
  \citep{2004ApJ...606L..17M,2005MNRAS.358..913V,2006MNRAS.368.1381L},
  through three body encounters \citep{2005MNRAS.358..913V}, as well
  as a variety of other processes.  Other sources of wandering BHs are
  the remnants of Population III stars
  \citep{2003MNRAS.340..647I,2005PhRvL..95a1301Z} as well as the the
  direct collapse of baryons into a BH
  \citep{2005PhRvD..72j3517B,2006MNRAS.368.1340M}.  In both of these
  cases, however, the BH either is presently located in the center of
  a galaxy or is ``naked'' without any stellar companions.  Such BHs
  are likely to only be located by their interaction with the matter
  that surrounds them
  \citep{2004MNRAS.354..629I,2004MNRAS.354..443I,2005MNRAS.358..913V,2005ApJ...628..873M,2005PhRvD..72j3517B}.
  Additionally, dwarf galaxies may be tidally stripped of their outer
  stars, leaving behind a BH surrounded by a massive cluster of stars
  similar to some globular clusters
  \citep{1988IAUS..126..603Z,1994ApJ...431..634B}.  In contrast, the
  new source of BHs presented in this paper uniquely have a population
  of bound stars that are only a fraction of the mass of the central BH.  }

In \S~\ref{sec:hier} we consider the formation history of the MW to
estimate the number and distribution of ejected BHs that may be found
in its halo today. We then follow the long term evolution of the star
clusters that are attached to them in \S~\ref{sec:relax}.  In
\S~\ref{sec:observe} we discuss the observational signatures of these
systems.  Finally, \S~\ref{sec:disc} summarizes our main results and
their implications.

\section{The Merger History of the MW and Ejected BHs in the Halo }
\label{sec:hier}

In the standard cosmological context of hierarchical galaxy formation
\citep{2008arXiv0809.0898S} the MW formed from an overdense region of
the universe with a present-day virial mass $M_{\rm vir} \approx
1$--$2\times 10^{12}\, \msun$
\citep{2002ApJ...573..597K,2008MNRAS.384.1459L}.  The threshold for
cooling by atomic transitions limits the minimum mass of star forming
galaxies at high redshifts to $M_{\rm gal}\sim 10^8\,\msun$
\citep{2006Natur.441..322W}.  If the earliest population of
star-forming galaxies had BHs at their center, then their shallow
potential would be insufficient to retain the recoiling BHs after
mergers with a BH mass ratio $q\equiv (M_{\bullet,1}/M_{\bullet,2})
\gtrsim 0.1$. Nevertheless, since most kicks are modest ($\lesssim
500\,\kms$), the ejected BHs remain confined to the MW.

We estimate the number of BHs in the MW halo, $N_{\rm BH}$, by
analyzing a large statistical ensemble of possible merger tree
histories for the MW using analytic models for the distribution of
kick velocities and galaxy profiles.  In particular, we generate
$\gtrsim 10^3$ Monte-Carlo realizations of the merger history of the
MW galaxy with the merger tree code made publicly available
by \citet{2008MNRAS.383..557P}. This code follows a modified form of
the extended \citet{1974ApJ...187..425P}  formalism and is normalized to
match the Millennium run numerical simulation
\citep{2005Natur.435..629S,2008MNRAS.383..546C}.  We assume that the
MW has a mass of $1.5\times 10^{12}\,\msun$, and restrict our
simulations to formation scenarios in which the MW undergoes no major
mergers ($q>1/3$) since a redshift $z=1$ \citep{2007ApJ...657..262D}.

We determine the properties of the halo from its circular velocity at
the virial radius \citep{2001PhR...349..125B},
\begin{equation}
  \label{eq:vcirc}
  v_{\rm c} = 24 \left( \frac{M_{\rm gal}}{10^8 h^{-1}\,\msun}
  \right)^{1/3} \left(\frac{\Omega_m}{\Omega_m^z}
    \frac{\Delta_c}{18\pi^2}\right)^{1/6} \left(\frac{1+z_{\rm
        merge}}{10}\right)^{1/2} \,\kms,
\end{equation}
where $\Delta_c=18\pi^2+82d-39d^2$, $d=\Omega_m^z-1$, $\Omega_m^z =
(\Omega_m (1+z)^3/ (\Omega_m (1+z)^3+\Omega_\Lambda)$, evaluated at
the merger redshift $z$ and we adopt the values for the
cosmological density parameters, $\Omega_m$ and $\Omega_\Lambda$, used
in the Millennium run \citep{2005Natur.435..629S}.
In each galaxy, we assume that the dark matter follows an NFW radial
profile \citep{1996ApJ...462..563N} with a concentration parameter
$c=4$ out to the virial radius, as expected for a newly formed
dark-matter halo \citep{2002ApJ...568...52W}.  In this case, the
escape velocity from the galaxy's center is $v_{\rm esc} \approx 2.8
v_{\rm c}$.  We scale the stellar velocity dispersion of the bulge in
the low-mass galaxies similarly to the MW and adopt $\sigma_\star =
v_{\rm c}/2$. The BH mass in the center of each galaxy is then
dictated by the observed $\msmbh$--$\sigma_\star$ relation
\citep{2002ApJ...574..740T}
\begin{equation}
  \label{eq:msigma}
\msmbh = 8.1\times 10^6 M_\odot 
\left(\frac{\sigma_\star}{100\,\kms}\right)^4 .
\end{equation}
We assume that the merger of the BHs is efficient, occurring before a
third BH is introduced into the system, and that the remnant BH mass
is $M_{\rm \bullet,final} = M_{\bullet,1}+M_ {\bullet,2}$. Although
the $\msmbh$--$\sigma_\star$ relation was initially determined from
the analysis of more massive BHs,
\citet{2005ApJ...619L.151B} and \citet{2006ApJ...641L..21G} have found that it
consistently extends to active galactic nuclei with BHs with masses
$\sim 10^5\,\msun$, {\new even without classical bulges \citep{2008ApJ...688..159G}.}

For each merger tree, we randomly assign a kick velocity from analytic
models of the kick velocity distribution of
\citet{2007ApJ...662L..63S} assuming that all the BHs have the same
spin with two different values: $a = 0.1$ and $0.9$.  A BH
is ejected from its host galaxy if the kick velocity is larger than
the escape velocity, $v_k > v_{\rm esc}$, and remains in the larger
potential of the MW if the final velocity is less than the escape
velocity of the MW halo, $v_{\rm ej} = \sqrt{v_{\rm k}^2 - v_{\rm
    esc}^2} \lesssim 500\,\kms$. Although the MW escape velocity
evolves by tens of percent as the MW halo assembles\footnote{The
  gravitational potential does not evolve during the linear growth of
  perturbations at redshifts $z\gtrsim 1$, and is only modestly
  enhanced in the final collapse of galaxy halos
  \citep{2006astro.ph..3360L}.}, we neglect this evolution in our
analysis.

In our realizations of the MW merger tree, there were an average of
$\approx 1500$ galaxy mergers for each run. Of these, $\approx 700$
were major mergers with $q > 0.1$. For BHs with spin $a = 0.9$
($a=0.1$) gravitational wave recoil ejected $\approx 570$ ($\approx
440$) BHs from their parent galaxy, and $N_{\rm BH} \approx 330$
($\approx 440$) remained within the MW halo. In all of our
simulations, the average kick velocity, even for the most massive BHs
ejected, is comparable to or less than the present day velocity
dispersion of the halo $\lesssim 200\,\kms$.  Thus, we expect that
most of the BHs trace the distribution of the dark matter in the halo.
{\new This is consistent with the results of
\citet{2006MNRAS.368.1381L} who used $N$-body simulations to follow
the kicked BHs in MW halo.  The BHs which were not ejected from
  their host galaxies settle back into the core of galaxies owing to
  dynamical friction, however, the vast majority of these galaxies
  have since merged to form the Milky Way.}

\begin{figure}
  \centering \includegraphics[width=\columnwidth]{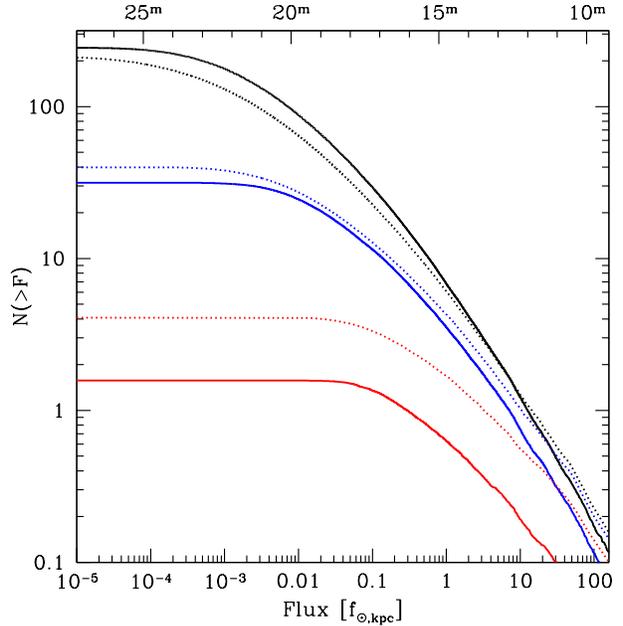}
  \caption{The cumulative distribution of ejected star clusters in the
    MW Halo.  Plotted is the flux distribution associated with BHs
    masses greater than $10^3\,\msun$ (black), $10^4\,\msun$ (blue),
    and $10^5\,\msun$ (red), in our models with BH spin $a=0.1$
    (solid) and $a=0.9$ (dashed) lines, plotted in units of the flux
    of the Sun at a distance of $1\,\kpc$ ($f_{\odot,{\rm kpc}}$).
    The top axis is labelled with the apparent bolometric magnitude of
    the clusters.  Nearly all BHs with $\msmbh \gtrsim 2\times
    10^3\,\msun$ have apparent magnitudes greater than 21, the rough
    magnitude limit of SDSS.  The mass distribution of the ejected BHs
    has approximately equal number per $\log\msmbh$ interval, with $\rmd
    N_{\rm BH}/ \rmd \msmbh \propto \msmbh^{-1}$.  }
  \label{fig:dist}
\end{figure}

In Figure~\ref{fig:dist}, we plot the cumulative flux distribution of
the remnant BHs, assuming that the BHs follow the present-day NFW
profile of the MW with $c = 12$ and $r_{\rm vir} = 200\,\kpc$
\citep{2002ApJ...573..597K,2008MNRAS.384.1459L}.  We determine the
luminosity from Eq.~(\ref{eq:nbound}) in \S~\ref{sec:relax}, and
report the flux in units of the flux from the Sun if it were at a
distance of $1\,\kpc$, $f_{\odot,{\rm kpc}}$, and also the apparent
bolometric magnitude (top axis).  Assuming that these clusters lose
little mass, we expect that nearly all remnant BHs with mass $\msmbh
\gtrsim 2\times 10^3\,\msun$ in the MW halo would be visible to the
depth of {\new {\it Sloan Digital Sky Survey} }\footnote{http://www.sdss.org} (SDSS).

In all our models, we find the mass distribution of the ejected BHs to
scale roughly as $\rmd N_{\rm BH} / \rmd \msmbh \propto \msmbh^{-1}$
(with the $a=0.1$ case showing a slightly steeper distribution on the
high mass end of BHs).  The most massive BH present in the halo is $
6.3^{+5.7}_{-2.8}\times 10^5\,\msun$ for $a=0.9$, and $
1.7^{+0.8}_{-0.5}\times 10^5\,\msun$ for $a=0.1$, where the quoted
uncertainty represents the one standard deviation. The closest BH that
has an apparent magnitude $< 21^{\rm m}$, the approximate depth of
SDSS, has an average mass of $\sim 3\times 10^3\,\msun$ and is $\sim
1\,\kpc$ away.  {\new For the typical distances and masses of the BHs}, dynamical friction is
negligible over the age of the universe.
{\new For example, the dynamical friction timescale of a $\sim 10^5\,\msun$
BH is $\gtrsim 10^{11}\,\yr$ at even $\sim 1\,$kpc from the galactic center \citep{1987gady.book.....B}. }

{\new Given the assumptions used here and the uncertainties in the
  lower limit for the BH mass, our results are consistent with the
  findings of previous analyses of BHs in the MW halo, even though
  only a few papers
  \citep[i.e.,][]{2005MNRAS.358..913V,2006MNRAS.368.1381L} explicitly
  account for GW recoil.  Previous studies that focused on BHs that
  formed in the smallest overdensities in the early universe ($\sim
  10^6\,\msun$ at $z\gtrsim 20$) found that there may be up to
  $10^3-10^4$ BHs with masses $\gtrsim 10^2\,\msun$
  \citep{2004MNRAS.354..427I,2005PhRvL..95a1301Z,2005PhRvD..72j3517B,2006MNRAS.368.1340M}.
  \citet{2005MNRAS.358..913V} and \citet{2006MNRAS.368.1381L} used
  more conservative estimates for the threshold of forming seed BHs,
  and accounted for the growth of the BH owing to accretion.  They
  concluded that there are more likely only $\sim 10^2$ such BHs in
  the halo.  Here, we sidestepped the issue of how and when the seed
  BHs formed, and assumed that all galaxies (with mass $\gtrsim
  10^8\,\msun$) have a central BH that follows the
  $\msmbh-\sigma_\star$ relationship, which consistently extends to
  the lowest mass BHs observed \citep{2006ApJ...641L..21G}.  With
  these conditions for the BH mass, our results are most consistent
  with \citet{2005MNRAS.358..913V} and \citet{2006MNRAS.368.1381L},
  however we do not follow the evolution of BHs with mass $\lesssim
  10^3\,\msun$.

}

\section{Structure and Evolution of Ejected Star Clusters}
\label{sec:relax}

{\new The number of stars that remain bound to the ejected BH will
  depend on the stellar distribution immediately before merger and the
  magnitude of the kick from GWs.  The stellar distribution is
  determined by the} relaxation timescale at the radius of influence of
the BH binary, $r_i = G \msmbh / \sigma_\star^2$, 
\citep{1976ApJ...209..214B}
\begin{equation}
  \label{eq:relaxtime} t_r = \frac{3(2\pi \sigma_\star^2)^{3/2}}{32 \pi^2 G^2
 m_\star^2 n_\star \ln{\Lambda}},
\end{equation}
where $\sigma_\star$ is the stellar velocity dispersion after the
galaxies merge, $m_\star$ is the average stellar mass, $n_\star$ is the number
density of stars at $r_i$, and $\ln \Lambda \approx \ln (\msmbh/M_\star)$
is the Coulomb logarithm.  From the $\msmbh$--$\sigma_\star$ relation
(Eq.~\ref{eq:msigma}) the relaxation timescale can be written as
\begin{equation}
  \label{eq:relaxsig}
  t_r \approx 10^{9} \left(\frac{\msmbh}{ 10^5 \,\msun}\right)^{5/4} \,\yr,
\end{equation}
if we assume that $m_\star = 1\,\msun$, and that the total mass in
stars interior of $r_i$ is $\sim 2\msmbh$. Within a relaxation
timescale, the stars form a density cusp within the radius of
influence around the central BH, with $n_\star(r) \propto
r^{-\alpha}$, where $\alpha = 1.75$ for a population of equal mass
stars and $\alpha \approx 1.5-2.0$ for varying mass distributions of
the stars and compact remnants \citep{1976ApJ...209..214B,
1977ApJ...216..883B, 2006ApJ...649...91F, 2006ApJ...645L.133H}.  If
the BHs coalesce on a time much shorter than the relaxation timescale
of the stars {\new owing to a gas rich merger}, then the stellar
density profile is likely much shallower with $\alpha \approx 1$ due
to the scattering off the inspiraling binary
\citep{2006ApJ...648..890M, 2007ApJ...671...53M}. However, new stars
might form out of the gas and introduce a new cusp of stars.  In
either case, however, the relaxation timescale of the stellar systems
is much shorter than a Hubble time, and the present-day distribution
of stars is different than its initial condition.

Assuming that the stars follow a power-law density profile and that
the total stellar mass within $r_i$ is $2\msmbh$, the stellar density
profile before the BH ejection is
\begin{equation}
  \label{eq:nstar}
  n_\star(r) = \frac{\msmbh}{m_\star}\frac{3-\alpha}{2\pi r_i^3}
  \left(\frac{r}{r_i} \right)^{-\alpha}.
\end{equation}
The kick is imparted to the BH merger remnant on a timescale much
shorter than the dynamical time of the star cluster.  In the frame of
the BH, all stars receive kicks with a reflex velocity $-v_k$.
Stars with total energies $\gtrsim -m_\star v_k^2/2$ will become unbound
to the BH. In the Keplerian potential of the BH, this roughly
corresponds to stars at $r \gtrsim r_k = \sqrt{G \msmbh/v_k^2} =
(\sigma_\star/v_k)^2 r_i$. The total number of stars that remain
bound the the BH is then,
\begin{equation}
  \label{eq:nbound}
  N_{\rm cl} \approx \frac{2 \msmbh}{m_\star}
  \left(\frac{v_k}{\sigma_\star}\right)^{2a - 6}.
\end{equation}
For $\alpha = 1.75$ the total number of stars that remain bound for a
minimally ejected BH with $v_k = 5.6\sigma_\star$ is then $\sim
4\times 10^{3} (\msmbh/10^5\msun)$. The actual number of stars may be
less due to the ejection of stars by the BH binary inspiral
\citep{2006ApJ...648..890M, 2007ApJ...671...53M}. These stars will
mainly remain within a radius $r_k \approx (\sigma_\star/v_k)^2 r_i
\ll 1\,\pc$, and have a $\alpha = 4$ profile for $r \gtrsim r_k$
\citep{2008ApJ...683L..21K}. Given the short relaxation timescale of
the star cluster and the lack of a source of new stars, the star
cluster will expand.

\begin{figure*}
  \centering
  \includegraphics[width=\columnwidth]{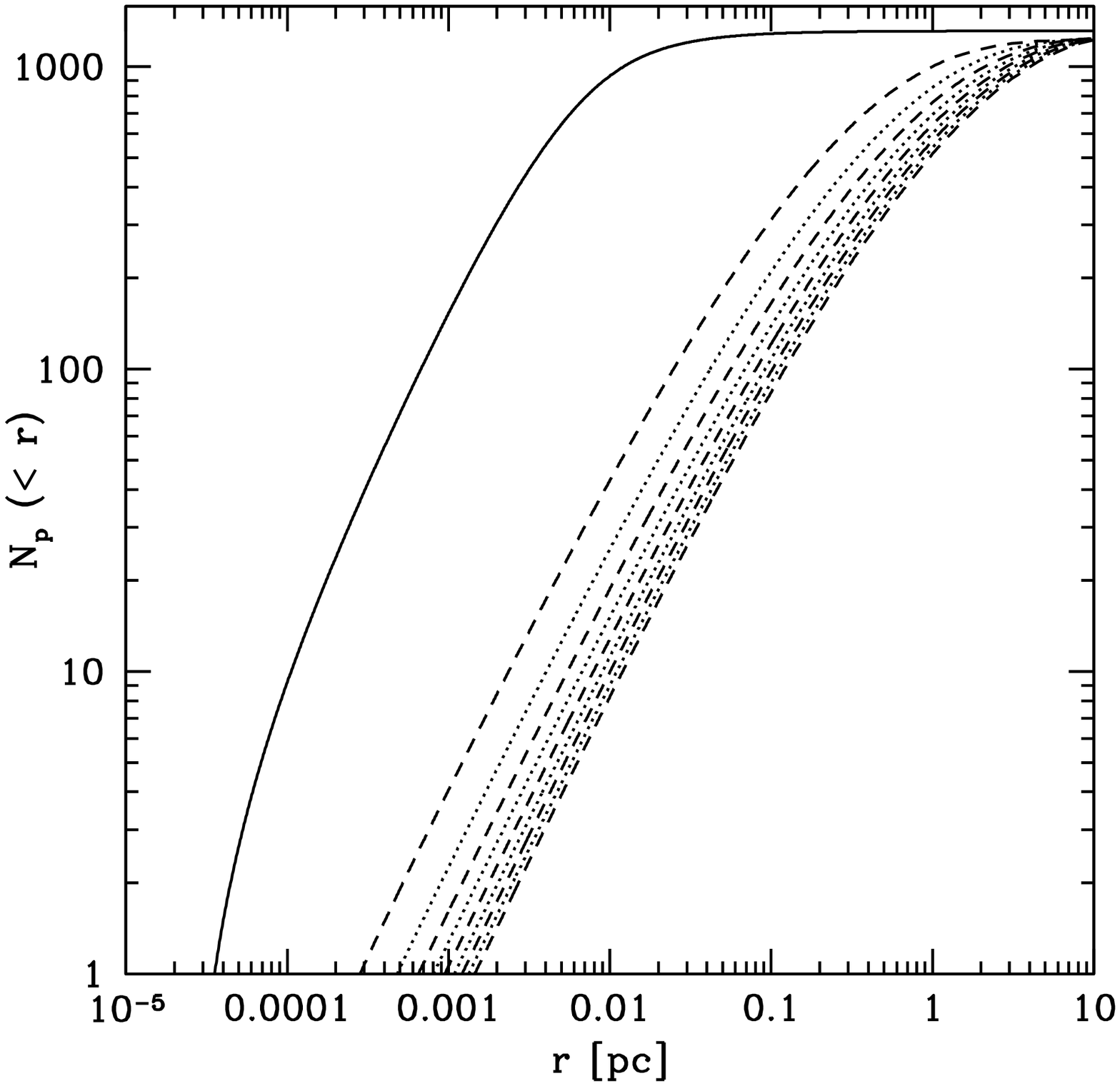}
  \includegraphics[width=\columnwidth]{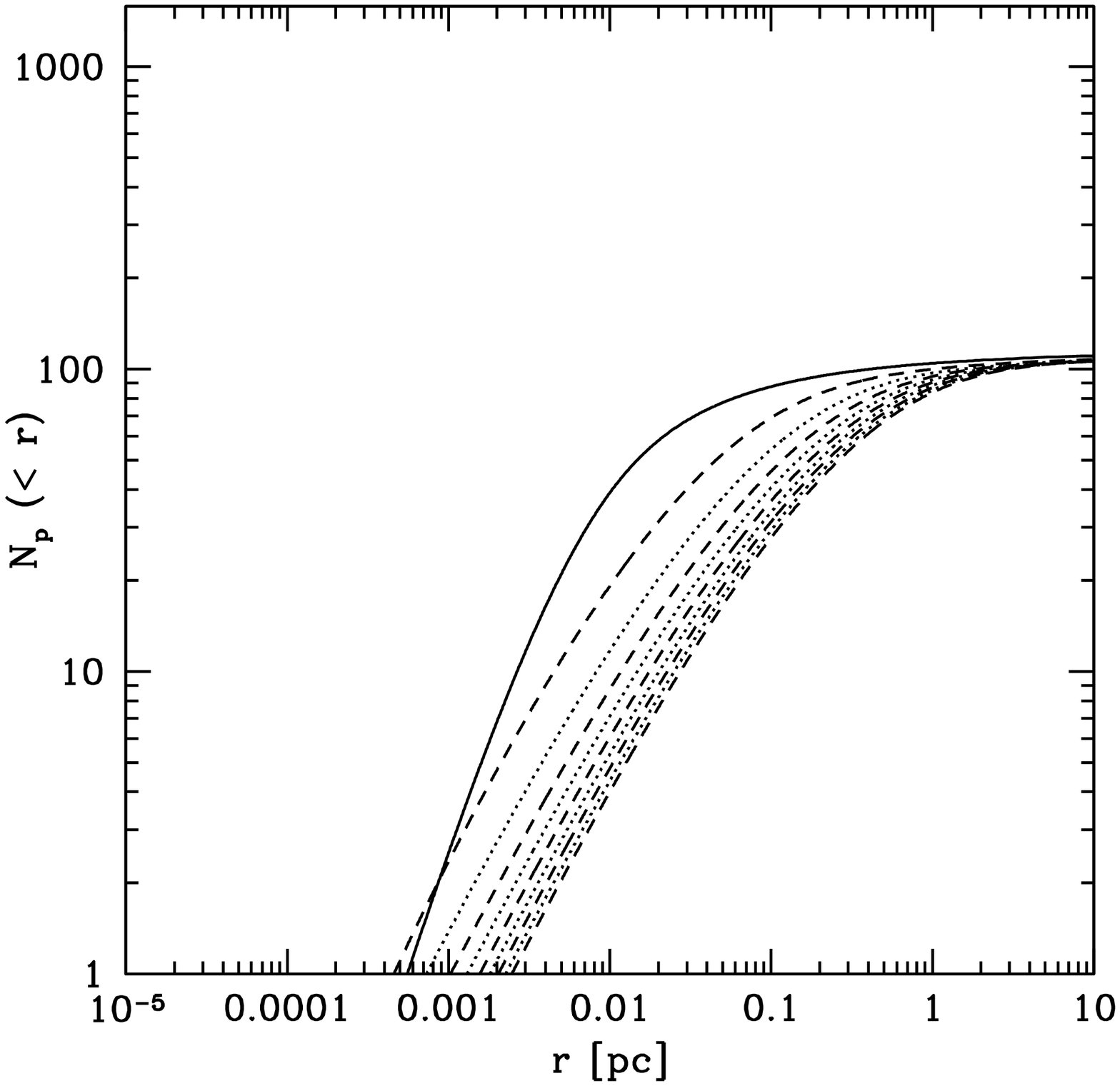}
  \caption{The total number of projected stars interior to $r$,
    $N_p(<r)$, for $\msmbh = 10^5\msun$ and $\alpha = 1.75$ (left)
    {\new and $\alpha = 1$ (right)}.  The solid line corresponds to
    the cluster immediately after being ejected from its parent galaxy
    with $v_k = 5.8 \sigma_\star$,{\new  normalized to have the same number
    density at the radius of influence}. The alternating dashed and
    dotted lines correspond to the projected number of stars after
    every $10 t_r \approx 650\,\Myr$.  Immediately after ejection, the
    cluster rapidly expands until its relaxation time becomes
    comparable to the age of the Universe, with very little mass
    loss. n For a $10^5\,\msun$ BH, the circular velocity of the stars
    is $\approx 66\,\kms (r/0.1\,\pc)^{-1/2}$.  }
  \label{fig:densityevolution}
\end{figure*}

We determine the long term evolution of the ejected cluster by
numerically solving the time dependent, angular momentum averaged,
Fokker-Planck equation for stars around a central massive object
\citep{1976ApJ...209..214B,1977ApJ...216..883B} {\new 
\begin{equation}
  \label{eq:fokkerplanck}
  \frac{\partial g(x,\tau)}{\partial \tau} = -x^{5/2}
  \frac{\partial}{\partial x} Q(x) - R(x),
\end{equation}
where $x = -E/(m_\star \sigma_\star^2)$ is the dimensionless energy,
$\tau = t / t_r$ is the dimensionless time, $g(x,\tau) = [(2 \pi
\sigma_\star^2)^{3/2} n_\star^{-1}]f(E)$ is the dimensionless
distribution function of the stars, $Q(x)$ is the rate stars flow to
energies larger than $x$, and $R(x)$ is the tidal disruption rate of
stars by the BH \citep{2008ApJ...683L..21K}. In these units,
\begin{eqnarray}
  \label{eq:flowrate}
  Q(x) = \int_{-\infinity}^{x_{\rm td}}
  \rmd y [{\rm max} (x,y)]^{-3/2} 
   \left(g(x) \frac{\partial
      g(y)}{\partial y} - g(y)\frac{\partial
      g(x)}{\partial x}\right).
\end{eqnarray}
and
\begin{equation}
  \label{eq:losscone}
  R(x) = \frac{g(x)^2}{ \ln[J_c(x)/J_{\rm LC}]},
\end{equation}
where $J_c(x) = G \msmbh \sigma_\star^{-1} (2 x)^{-1/2}$ is the
angular momentum of a circular orbit at a radius $r \approx r_i / (2 x)$,
and $J_{\rm LC}$ is the maximum angular momentum for the star to be
destroyed or consumed by the BH. For a star of radius $R_\star$ $J_{\rm
  LC} = G \msmbh \sigma^{-1}x_{\rm td}^{-1} (2(x+x_{\rm
  td}))^{1/2}$, where $x_{\rm td} \approx (\msmbh / m_\star)^{-1/3} r_i/R_\star$.
We assume that far from the BH the stars initially follow a Maxwellian
distribution with dispersion $\sigma_\star$. In dimensionless units,
the distribution function of the unbound stars is just $g(x)=exp(x)$
for $x>0$.  For a power-law distribution function with $g(x)\propto
x^{p}$, the density profile of the stars will also follow a power-law
 $n_\star \propto r^{-\alpha}$, with $\alpha = -p - 3/2$. }  

We have performed two sets of calculations, {\new first for a binary
  that merges because of stellar relaxation, and then for a gas rich
  merger. For the first case}, we evolved the system for one
relaxation timescale, at which point the system has reached steady
state, and then we removed all unbound stars by setting $g(x) = 0$ for
$x>0$.  The initial kick on the system breaks the spherical symmetry
assumed here. However, after a relaxation timescale we expect the
system to reach spherical symmetry again.  To represent the kick on
the bound stars, we scale the distribution function of stars as $g(x)
\rightarrow g(x) z^{2.5}/(1+z^{2.5})$, where $z = x /
(v_k/\sigma_\star)^2$. This yields an asymptotic density profile with
$n \propto r^{-4}$ for $r \gtrsim r_k$, as expected immediately after
the kick \citep{2008ApJ...683L..21K}.  To examine the evolution of the
star system if the BHs coalesce through a gas rich merger
\citep{2008arXiv0808.3088C}, we adopt $g(x) \propto x^{-1/2}
z^{3}/(1+z^{3})$, normalized by the same number density and relaxation
timescale used in Eq.~(\ref{eq:relaxtime}). {\new This yields $\alpha
  \sim 1$ for $r \lesssim r_k$ and $\alpha \sim 4$ for $r \gtrsim
  r_k$, as expected following a merger
  \citep{2006ApJ...648..890M,2007ApJ...671...53M}.}  This run results
in $\sim 10\%$ the number of bound stars as with $\alpha = 1.75$. Its
importance, however, is uncertain given the short relaxation
timescales for these systems.

Shortly after the GW recoil kicks the BH, the star cluster expands as
a result of two-body relaxation and evolves until the relaxation
timescale of the cluster becomes comparable to the age of the
system. Figure~\ref{fig:densityevolution} shows the evolution of the
projected number of stars interior to a radius $r$ for a BH with mass
$10^5\,\msun$ for both $\alpha = 1.75$ and $\alpha = 1$ density
profiles. Since our simulations indicate little mass loss, the
condition that the relaxation timescale of the system
(Eq.~\ref{eq:relaxtime}) be comparable to the age of the Universe
implies that the cluster should expand by a factor of order
$(t_h/t_r)^{1/3}\sim 10$.  Even then, the present day cluster size is
less than the initial radius of influence of the BH, $r_i \lesssim
1\,\pc$, and much smaller than the tidal radius of the system,
distinguishing them from globular clusters with a similar stellar mass
(owing to the gravitational binding provided by the central BH).

In our simulations there is very little mass loss due to the tidal
disruption of stars or the ejection of stars from the system.  Since
the cluster expands so rapidly, the rate of tidal disruption of stars
rapidly decreases, and is much lower than the rate associated with a
comparable BH in the nucleus of a galaxy. 

Our simulations only treat the evolution of the cluster due to many
small-angle scatterings (two-body relaxation), and does not include
the strong scattering that might occur in a dense stellar
environment. Such scatterings would launch stars on eccentric orbits
that take them outside of the cluster to about the tidal radius, thus
enhancing the mass loss from the system. The existence of a halo of
such stars provides an additional signature unique to the compact star
clusters associated with recoiled BHs. {\new We also ignore the
  effects of resonant relaxaation, which can deplete the number of
  stars both before and after the ejection of of the BH
  \citep{1996NewA....1..149R,1998MNRAS.299.1231R}. Although resonant
  relaxation has been approximately accounted for in
  Eq.~\ref{eq:fokkerplanck} by averaging over angular momentum
  \citep{2006ApJ...645.1152H}, a full multidimensional analysis
  accounting for angular momemtum is best suited for these clusters.}

\section{Observing Ejected Systems}
\label{sec:observe}

\noindent {\em Size and Structure:} Based on the results in
\S~\ref{sec:hier}, we estimate that there should be $\sim 300$
recoiled BHs in the MW halo. Equation~(\ref{eq:nbound}) implies that
these BHs are surrounded by $\sim 4\times 10^{3} (\msmbh/10^5
m_\star)$ stars with a half mass radius, $r_h \lesssim 1\,\pc$. At the
typical distance to the most massive recoiled BH, $\sim 100\,\kpc$,
the cluster will have a typical angular size $\lesssim 1$ arcsec,
below the resolution limit of most existing sky surveys.  Thus, the
cluster will appear as a point source and might be confused as a
single foreground star with unusual colors.  The closest clusters are
$\sim 1\,\kpc$ away and may be a few arcmin in size.  However, these
clusters surround the smallest BHs ($\sim 10^3\,\msun$) with the
fewest number of stars, and their present day distribution is
uncertain owing to strong encounters between the stars in the cluster
as well as stars in the disk.  To date, no star cluster has been
identified in the MW with such a small size; the smallest star cluster
discovered has a size of $\sim 3\,\pc$
\citep{2007ApJ...669..337K}. Clusters at distances of tens of kpc
could in principle be distinguished from point sources by comparing
their extended image to the point-spread-function of the telescope.
{\new In comparison, the stellar mass of globular clusters and the tidally
  stripped cores of dwarf galaxies completely outweights the total
  mass of any central BH in such systems.  By measuring the total mass of
  stars one can approximately determine how large the kick velocity
  may have been.}

\noindent {\em Color:} A recoiled star cluster can be distinguished from
single foreground stars by its anomalous color.  The cluster is
likely old, $\sim 1-10\,\Gyr$, and to first order should have colors
similar to globular clusters of a similar age and metallicity.

\noindent {\em Spectra:} The velocity dispersion profile of the
ejected star clusters should be Keplerian. The unusually large width
of their spectral lines (greather than one hundred km~s$^{-1}$) can be
used to distinguish these clusters from foreground stars, which
possess much lower rotational velocities.

The compact star clusters around ejected BHs are distinguishable from
cores of globular clusters by their internal (Keplerian) velocity
dispersion of order one hundred km~s$^{-1}$ and their high
mass-to-light ratio owing to the central BH.

\noindent {\em Proper Motion:} The most massive systems that reside near the
virial radius of the galaxy and have a relatively small proper motion.
Nevertheless, future surveys like
GAIA\footnote{http://www.esa.int/esaSC/},
Pan-STARRS\footnote{http://pan-starrs.ifa.hawaii.edu/}, and
LSST\footnote{http://www.lsst.org/}, could aim to
measure the proper motion of such clusters, and be used in conjunction
with expected cluster colors to distinguish the clusters from
foreground stars. The expected proper motion of these clusters is
$\sim 10^{-3}$ arcsec~yr$^{-1}$.

\noindent {\em The Local Group:} Given the constrained distances to M31 and M33,
color-magnitude diagrams could be used to identify candidate star
clusters around ejected BHs, and spectroscopic follow up could reveal
the high dispersions expected from the stars very near the BH. At the
distance of M31, a cluster with $r_{\rm h} \sim 0.5\,\pc$ would have
an angular size of $\sim 0.1$\,arcsec, and could in principle be
resolved by HST\footnote{http://www.stsci.edu/hst/}, JWST\footnote{http://www.stsci.edu/jwst/},  or ground-based telescopes with
adaptive optics.  

{\new 
  \noindent {\em Gas Accretion:} A faint radio glow from the central
  black hole may also be detectable as it accretes material from
  stellar winds in the surrounding cluster of stars
  \citep{2004MNRAS.350..725L}.  Rarely, these BHs may accrete
  gas from the interstellar medium or molecular clouds as they pass
  through the disk of the galaxy
  \citep{1998ApJ...495L..85F,2003MNRAS.340..647I,2005ApJ...628..873M,2005MNRAS.358..913V,2006MNRAS.368.1340M}, however, for the number
  of BHs expected in the MW, such a scenario is unlikely to be observed
  \citep{2005MNRAS.358..913V}.
}

\noindent {\em Dark Matter Annihilation:} The recoiled BHs may be surrounded by a
dense cusp of dark matter, which was adiabatically compressed by the
baryonic condensation and BH growth in its parent low-mass
galaxy. These high density cusps can dramatically increase the rate of
dark matter annihilation, and may be visible through the relativistic
by products of the annihilation
\citep{2005PhRvL..95a1301Z,2005PhRvD..72j3517B, 2008arXiv0809.0894S}.

\section{Summary and Discussion}
\label{sec:disc}

Based on a large statistical ensemble of merger tree histories for the
MW, we have found that hundreds of GW recoiled BHs should reside
within the MW halo today. The BHs have a mass $\msmbh \gtrsim
10^3\,\msun$, and should be surrounded by a cluster of stars that were
tightly bound to the BH when it was ejected.  The most massive BH
weighs $\sim 1-6 \times 10^5\,\msun$, and is surrounded by a compact star
cluster $\lesssim 1\,$pc in size with a stellar velocity dispersion of
hundreds of km~s$^{-1}$.  High-resolution adaptive optics imaging
along with spectroscopy of the star clusters can constrain the
distance and mass of the BH in the center of the cluster based on
orbits of individual stars for the closest clusters \citep{2005ApJ...620..744G,
  2005ApJ...628..246E}.

The number of recoiled BHs in the MW is most sensitive to the fraction
of low mass galaxies that harbored BHs in their centers, as well as to
the merger history of such galaxies during the formation of the MW. We
emphasize that most of these galactic ``building blocks'' were made at
high redshifts and were disrupted during the assembly process of the
MW; hence, they likely had different properties than the remaining
dwarf satellites of the MW today.  The future discovery of the
population of ejected star clusters will provide a unique probe of the
early history of the Milky Way, as well as the distribution and
evolution of low mass BHs. It would also open a new window to
exploring the low-mass end of the population of nuclear BHs in
high-redshifts galaxies.  \citet{2008arXiv0809.5046M}, whose work we
learned of after submission of this manuscript, arrived at a similar
conclusion independently, with a slightly different application to the
nearby Virgo cluster.

\section*{Acknowledgements}
We thank the GALFORM team for making their merger tree code
available. This work is supported in part by NASA grant NNX08AL43G, by
FQXi, and by Harvard University funds.

\bibliography{p}

\end{document}